\documentclass[aps,prb,floatfix,showpacs,preprint]{revtex4}

\usepackage{graphicx}

\begin{document}

\title{Alternative perturbation approaches in classical mechanics}

\author{P. Amore}\email{paolo@cgic.ucol.mx}

\author{A. Raya}\email{raya@ucol.mx}

\affiliation{Facultad de Ciencias, Universidad de Colima, \\
Bernal D\'{\i}az del Castillo 340, Colima, Colima, Mexico}

\author{Francisco M. Fern\'{a}ndez}\email{fernande@quimica.unlp.edu}
\affiliation{INIFTA (Conicet,UNLP), Diag. 113 y 64 S/N, \\
Sucursal 4, Casilla de Correo 16, 1900 La Plata, Argentina}

\date{\today}

\begin{abstract}
We discuss two alternative methods, based on the Lindstedt--Poincar\'{e}
technique, for the removal of secular terms from the equations of
perturbation theory. We calculate the period of an anharmonic oscillator by
means of both approaches and show that one of them is more accurate for all
values of the coupling constant.
\end{abstract}


\maketitle

\section{\label{sec:intro}Introduction}

Straightforward application of perturbation theory to periodic nonlinear
motion gives rise to secular terms that increase in time in spite of the
fact that the trajectory of the motion is known to be bounded \cite{N81,F00}%
. One of the approaches commonly used to remove those unwanted secular terms
is the method of Lindstedt--Poincar\'{e} \cite{N81,F00}, recently improved
by Amore \emph{et al.} \cite{AA03,AL04} by means of the delta expansion and
the principle of minimal sensitivity.

There is also another technique that resembles the method of
Lindstedt--Poincar\'{e} which is suitable for the removal of secular terms
\cite{M70}. Discussion and comparison of such alternative approaches may be
most fruitful for teaching perturbation theory in advanced undergraduate
courses on classical mechanics.

In Sec.~\ref{sec:Model} we present a simple nonlinear model to which we
apply the alternative perturbation approaches in subsequent sections. In
Sec.~\ref{sec:secular} we apply straightforward perturbation theory and
illustrate the outcome of secular terms. In Sec.~\ref{sec:L-P} we show how
to remove those secular terms by means of the Lindstedt--Poincar\'{e} method
\cite{N81,F00}. In Sec.~\ref{sec:ALP} we develop an alternative method that
appears in another textbook \cite{M70} and that closely resembles the method
of Lindstedt--Poincar\'{e}. In Sec. \ref{sec:VLP} we describe an improvement
to the method of Lindstedt--Poincar\'{e} proposed by Amore \emph{et al.}
\cite{AA03,AL04}. Finally, in Sec.~\ref{sec:results} we compare the period
of the motion calculated by all those approaches.

\section{\label{sec:Model} The Model}

In order to discuss and compare the alternative perturbation approaches
mentioned above, we consider the simple, nonlinear equation of motion
\begin{equation}
\ddot{x}(t)+x(t)=-\lambda x^{3}(t)  \label{eq:Newton}
\end{equation}
with the initial conditions $x(0)=1$ and $\dot{x}(0)=0$. In Appendix~\ref
{sec:Appendix_A} we show that we can derive this differential equation from
the equation of motion for a particle of mass $m$ in a polynomial anharmonic
potential with arbitrary quadratic and quartic terms. Notice that $E=\dot{x}%
^{2}/2+x^{2}/2+\lambda x^{4}/4=1/2+\lambda /4$ is an integral of the motion
for~(\ref{eq:Newton}) and that the motion is periodic for all $\lambda >-1$
for the initial conditions indicated above.

\section{\label{sec:secular} Secular Perturbation Theory}

The straightforward expansion of $x(t)$ in powers of $\lambda $
\begin{equation}
x(t)=\sum_{j=0}^{\infty }x_{j}(t)\lambda ^{j}
\end{equation}
leads to the perturbation equations
\begin{eqnarray}
\ddot{x}_{0}(t)+x_{0}(t) &=&0  \nonumber \\
\ddot{x}_{1}(t)+x_{1}(t) &=&-x_{0}^{3}(t)  \nonumber \\
\ddot{x}_{n}(t)+x_{n}(t)
&=&-\sum_{j=0}^{n-1}\sum_{k=0}^{j}x_{k}(t)x_{j-k}(t)x_{n-j-1}(t),\;n=2,3,%
\ldots
\end{eqnarray}
with the boundary conditions $x_{j}(0)=\delta _{j0}$ and $\dot{x}_{j}(0)=0$
for all $j\geq 0$. Clearly, the solution of order zero is $x_{0}(t)=\cos (t)$%
.

All those perturbation equations are of the form $\ddot{y}(t)+y(t)=f(t)$,
where $f(t)$ is a linear combination of $\cos (jt)$, $j=1,2,\ldots $. Such
differential equations, which are commonly discussed in introductory
calculus courses, are in fact suitable for illustrating the advantage of
using available computer algebra systems.

It is well known that any term proportional to $\cos (t)$ in $f(t)$ gives
rise to a secular term after integration \cite{N81,F00}. For example, since $%
x_{0}^{3}(t)=[\cos (3t)+3\cos (t)]/4$ we obtain
\begin{equation}
x_{1}(t)=\frac{1}{8}\left[ \frac{\cos (3t)-\cos (t)}{4}-3t\sin (t)\right]
\end{equation}
that clearly shows that $|x_{1}(t)|$ grows unboundedly with time in spite of
the fact that the exact motion is periodic for all $\lambda >-1$.

\section{\label{sec:L-P} Method of Lindstedt--Poincar\'{e}}

There are several suitable mathematical techniques that overcome the problem
of secular terms mentioned above \cite{N81}. For example, the method of
Lindstedt--Poincar\'{e} is based on the change of the time variable
\begin{equation}
s=\sqrt{\gamma }t  \label{eq:s(t)_LP}
\end{equation}
where $\sqrt{\gamma }$ plays the role of the frequency of the motion, and,
therefore, the period $T$ is given by
\begin{equation}
T=\frac{2\pi }{\sqrt{\gamma }}.  \label{eq:T_LP}
\end{equation}
The equation of motion~(\ref{eq:Newton}) thus becomes
\begin{equation}
\gamma x^{\prime \prime }(s)+x(s)=-\lambda x^{3}(s)
\end{equation}
where the prime stands for differentiation with respect to $s$. If we expand
both $x(s)$ and $\gamma $ in powers of $\lambda $
\begin{equation}
\gamma =\sum_{j=0}^{\infty }\gamma _{j}\lambda ^{j}
\end{equation}
with $\gamma _{0}=1$, then we obtain the set of equations
\begin{eqnarray}
x_{0}^{\prime \prime }(s)+x_{0}(s) &=&0  \nonumber \\
x_{1}^{\prime \prime }(s)+x_{1}(s) &=&-x_{0}^{3}(s)-\gamma _{1}x_{0}^{\prime
\prime }(s)  \nonumber \\
x_{n}^{\prime \prime }(s)+x_{n}(s)
&=&-\sum_{j=0}^{n-1}\sum_{k=0}^{j}x_{k}(s)x_{j-k}(s)x_{n-j-1}(s)-%
\sum_{j=1}^{n}\gamma _{j}x_{n-j}^{\prime \prime }(s),\;  \nonumber \\
n &=&2,3,\ldots
\end{eqnarray}
We choose the value of the coefficient $\gamma _{n}$ in order to
remove the secular term from the perturbation equation of order
$n$. For example, it follows from $-x_{0}^{3}-\gamma
_{1}x_{0}^{\prime \prime }=\left( \gamma _{1}-3/4\right) \cos
(t)-\cos (3t/4)$ that $\gamma _{1}=3/4$ is the right choice at
first order. Proceeding exactly in the same way at higher orders
we obtain the coefficients
\begin{equation}
\gamma _{1}=\frac{3}{4}\ \ ,\ \ \gamma _{2}=-\frac{3}{128}\ \ ,\ \ \gamma
_{3}=\frac{9}{512}  \label{eq:gammaj_LP}
\end{equation}
and the approximate period
\begin{equation}
T=\frac{32\sqrt{2}\pi }{\sqrt{\left( 9\lambda ^{3}-12\lambda ^{2}+384\lambda
+512\right) }}  \label{eq:T_LP3}
\end{equation}

\section{\label{sec:ALP} Alternative Lindstedt--Poincar\'{e} Technique}

Perturbation theory provides a $\lambda $--power series for the frequency of
the motion $\omega $; for example, for our model it reads
\begin{equation}
\omega ^{2}=1+w_{1}\lambda +w_{2}\lambda ^{2}+\ldots  \label{eq:omega^2_PT}
\end{equation}
An alternative perturbation approach free from secular terms is based on the
substitution of this expansion into the equation of motion~(\ref{eq:Newton})
followed by an expansion of the resulting equation
\begin{equation}
\ddot{x}(t)+\left( \omega ^{2}-w_{1}\lambda -w_{2}\lambda ^{2}+\ldots
\right) x(t)=-\lambda x^{3}(t)
\end{equation}
in powers of $\lambda $, as if $\omega $ were independent of the
perturbation parameter \cite{M70}. The perturbation equations thus produced
read
\begin{eqnarray}
\ddot{x}_{0}(t)+\omega ^{2}x_{0}(t) &=&0  \nonumber \\
\ddot{x}_{1}(t)+\omega ^{2}x_{1}(t) &=&-x_{0}^{3}(t)+w_{1}x_{0}(t)  \nonumber
\\
\ddot{x}_{n}(t)+\omega ^{2}x_{n}(t)
&=&-\sum_{j=0}^{n-1}\sum_{k=0}^{j}x_{k}(t)x_{j-k}(t)x_{n-j-1}(t)-%
\sum_{j=1}^{n}w_{j}x_{n-j}(t),\;  \nonumber \\
n &=&2,3,\ldots .
\end{eqnarray}
Notice that $x_{0}=\cos (\omega t)$ depends on $\omega $ and so does each
coefficient $w_{n}$ that we set to remove the secular term at order $n$.
Consequently, we have to solve the partial sums arising from truncation of
the series~(\ref{eq:omega^2_PT}) for $\omega $ in order to obtain the
frequency and the period
\begin{equation}
T=\frac{2\pi }{\omega }.  \label{eq:T_Marion}
\end{equation}
in terms of $\lambda $ \cite{M70}.

A straightforward calculation through third order yields
\begin{eqnarray}
w_{1} &=&\frac{3}{4} \ \ , \ \ w_{2} = -\frac{3}{128\omega ^{2}} \ \ , \ \
w_{3} = 0  \label{eq:wj's}
\end{eqnarray}
from which we obtain
\begin{equation}
\omega =\frac{\sqrt{\left( \sqrt{30\lambda ^{2}+96\lambda +64}+2(3\lambda
+4)\right) }}{4}  \label{eq:omega_Marion}
\end{equation}

\section{\label{sec:VLP} Variational Lindstedt--Poincar\'{e}}

Amore \emph{et al.} \cite{AA03,AL04} have recently proposed a variational
method for improving the Lindstedt--Poincar\'{e} technique. It consists of
rewriting equation~(\ref{eq:Newton}) as
\begin{equation}
\ddot{x}(t)+\left( 1+\alpha ^{2}\right) x(t)=\delta \left( -\lambda
x^{3}(t)+\alpha ^{2}x(t)\right)   \label{eq:eq_mot_varLP}
\end{equation}
where $\alpha $ is an adjustable variational parameter, and $\delta $ is a
dummy perturbation parameter that we set equal to unity at the end of the
calculation. When $\delta =1$ the modified equation of motion~(\ref
{eq:eq_mot_varLP}) reduces to equation (\ref{eq:Newton}) that is independent
of $\alpha $. Following the Lindstedt--Poincar\'{e} technique we change the
time variable according to equation~(\ref{eq:s(t)_LP}) thus obtaining
\begin{equation}
\gamma x^{\prime \prime }(s)+(1+\alpha ^{2})x(s)=\delta \left( -\lambda
x^{3}(s)+\alpha ^{2}x(s)\right) .
\end{equation}
We then expand both $x$ and $\gamma $ in powers of $\delta $ and proceed
exactly as is Sec.~\ref{sec:L-P}, except that in this case $\gamma
_{0}=1+\alpha ^{2}$. Thus we obtain
\begin{eqnarray}
x_{0}^{\prime \prime }(s)+x_{0}(s) &=&0  \nonumber \\
x_{1}^{\prime \prime }(s)+x_{1}(s) &=&\frac{1}{\gamma _{0}}\left( -\lambda
x_{0}^{3}(s)-\gamma _{1}x_{0}^{\prime \prime }(s)+\alpha ^{2}x_{0}(s)\right)
\nonumber \\
x_{n}^{\prime \prime }(s)+x_{n}(s) &=&\frac{1}{\gamma _{0}}\left(
-\sum_{j=0}^{n-1}\sum_{k=0}^{j}x_{k}(s)x_{j-k}(s)x_{n-j-1}(s)\right. \;
\nonumber \\
&&\left. -\sum_{j=1}^{n}\gamma _{j}x_{n-j}^{\prime \prime
}(s)+\alpha ^{2}x_{n-1}(s)\right) ,\;n =2,3,\ldots .
\end{eqnarray}
The period of the motion is given by equation~(\ref{eq:T_LP}).

Choosing the value of $\gamma _{n}$ in order to remove the secular term from
the perturbation equation of order $n$ we obtain
\begin{equation}
\gamma _{1}=\frac{3\lambda -4\alpha ^{2}}{4}\ \ ,\ \ \gamma _{2}=-\frac{%
3\lambda ^{2}}{128\left( 1+\alpha ^{2}\right) }\ \ ,\ \ \gamma _{3}=\frac{%
3\lambda ^{2}\left( 3\lambda -4\alpha ^{2}\right) }{512\left( 1+\alpha
^{2}\right) ^{2}}.  \label{eq:gammaj_var}
\end{equation}
Notice that these coefficients reduce to those in
equation~(\ref{eq:gammaj_LP}) (multiplied by the proper power of
$\lambda $) when $\alpha =0$. Since the actual value of $\gamma $
is independent of $\alpha $ when $\delta =1$, we make use of the
principle of minimal sensitivity developed in Appendix~\ref
{sec:Appendix_B}. The root of
\begin{equation}
\frac{d}{d\alpha }\sum_{j=0}^{3}\gamma _{j}=0  \label{eq:PMS}
\end{equation}
is
\begin{equation}
\alpha =\frac{\sqrt{3\lambda }}{2}  \label{eq:alpha_PMS}
\end{equation}
and we thus obtain
\begin{equation}
T=\frac{8\sqrt{2}\pi (3\lambda +4)}{\sqrt{\left( 207\lambda ^{3}+852\lambda
^{2}+1152\lambda +512\right) }}.  \label{eq:T_LP_PMS1}
\end{equation}

For this particular problem we find that the value of $\alpha $ given by the
PMS condition~(\ref{eq:alpha_PMS}) is such that $\gamma _{2j+1}=0$ for all $%
j\geq 0$.

\section{\label{sec:results} Results and Discussion}

\begin{figure}[H]
\begin{center}
\includegraphics[width=10cm]{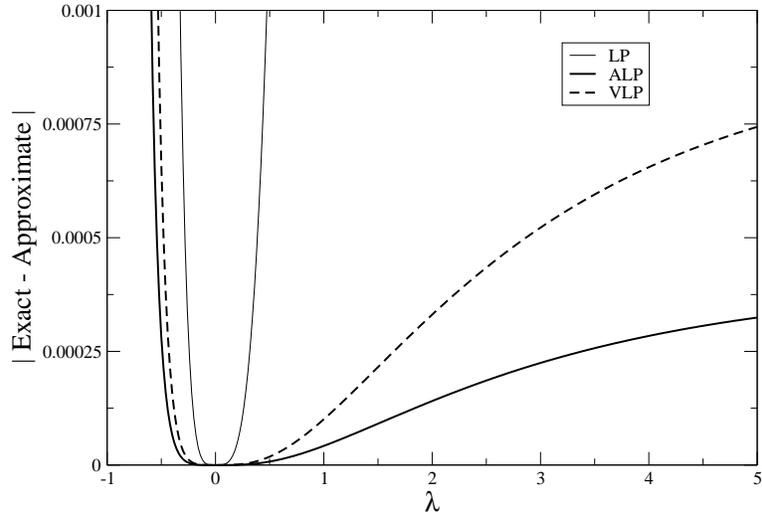}
\end{center}
\caption{Absolute error on the period $\left| T_{exact}-T_{approx} \right|$
as a function of the coupling $\lambda$ for the Lindstedt-Poincar\'e (LP,
solid thin line), the Alternative Lindstedt-Poincar\'e (ALP, solid bold
line) and the Variational Lindstedt-Poincar\'e (VLP, dashed line) methods.}
\label{Fig_1}
\end{figure}

Fig.~\ref{Fig_1} shows the period as a function of $\lambda $
given by the perturbation approaches discussed above and by the
exact expression \cite{N81}
\begin{equation}
T=\frac{2}{\sqrt{1+\lambda }}\int_{0}^{\pi }\frac{d\theta }{\sqrt{1-\frac{%
\lambda \sin (\theta )^{2}}{2(1+\lambda )}}}.  \label{eq:T_exact}
\end{equation}
We appreciate that the variational Lindstedt--Poincar\'{e} method \cite
{AA03,AL04} yields more accurate results than the straightforward
Lindstedt--Poincar\'{e} technique \cite{N81,F00}, and that the alternative
Lindstedt--Poincar\'{e} approach \cite{M70} is the best approach, at least
at third order of perturbation theory.

All those expressions yield the correct value $T(0)=2\pi $ and become less
accurate as $\lambda $ increases. However, two of them give reasonable
results even in the limit $\lambda \rightarrow \infty $. The exact value is
\begin{equation}
\lim_{\lambda \rightarrow \infty }\sqrt{\lambda }T^{exact}=2\sqrt{2}%
\int_{0}^{\pi }\frac{d\theta }{\sqrt{1+\cos (\theta )^{2}}}=7.4162987.
\end{equation}
The standard Lindstedt--Poincar\'{e} technique fails completely as shown by
\begin{equation}
\lim_{\lambda \rightarrow \infty }\sqrt{\lambda }T^{LP}=0.
\end{equation}
The variational improvement proposed by Amore \emph{et al.} \cite{AA03,AL04}
corrects this anomalous behavior
\begin{equation}
\lim_{\lambda \rightarrow \infty }\sqrt{\lambda }T^{VLP}=\frac{8\sqrt{46}\pi
}{23}=7.4112410.
\end{equation}
Finally, the alternative Lindstedt--Poincar\'{e} method \cite{M70} gives the
closest approach
\begin{equation}
\lim_{\lambda \rightarrow \infty }\sqrt{\lambda }T^{ALP}=\pi \sqrt{\left( 64-%
\frac{32\sqrt{30}}{3}\right) }=7.4185905.
\end{equation}

\appendix

\section{\label{sec:Appendix_A} Dimensionless equations}

Transforming an equation of physics into a dimensionless mathematical
equation is most convenient for at least two reasons. First, the latter is
much simpler and reveals more clearly how it can be solved. Second, the
dimensionless equation exhibits the actual dependence of the solution on the
parameters of the physical model.

In order to illustrate how to convert a given equation into a dimensionless
one we consider a particle of mass $m$ moving in the potential
\begin{equation}
V(q)=\frac{v_{2}}{2}q^{2}+\frac{v_{4}}{4}q^{4}.
\end{equation}
The equation of motion is
\begin{equation}
m\ddot{q}=-v_{2}q-v_{4}q^{3}
\end{equation}
and we assume that $q(0)=q_{0}$ and $\dot{q}(0)=v_{0}$.

We define a new independent variable $s=\omega _{0}t+\phi $, where $\phi $
is a phase, and $\omega _{0}=\sqrt{v_{2}/m}$ is the frequency of the motion
when $v_{4}=0$. Suppose that $\dot{q}=0$ and $q=A$ at $t=t_{1}$; then we
define the dependent variable $x(s)=q(t)/A$ and choose $\phi =-\omega
_{0}t_{1}$ so that $x(s)$ is a solution of the differential equation
\begin{equation}
x^{\prime \prime }(s)+x(s)=-\lambda x(s)^{3}  \label{eq:dif_q}
\end{equation}
where $\lambda =v_{4}A^{2}/v_{2}$ and the initial conditions become $x(0)=1$
and $x^{\prime }(0)=0$.

The dimensionless differential equation~(\ref{eq:dif_q}) resembles the
equation of motion for a particle of unit mass moving in the potential $%
V(x)=x^{2}/2+\lambda x^{4}/4$. Its period $T$ depends on $\lambda $ and,
therefore, the expression for the period of the original problem $T^{\prime
}=T/\omega _{0}$ clearly reveals the way it depends upon the model
parameters $m$, $v_{2}$, $v_{4}$, and $A$.

\section{\label{sec:Appendix_B}Variational perturbation theory}

Variational perturbation theory is a well--known technique for obtaining an
approximation to a property $P(\lambda )$ in a wide range of values of the
parameter $\lambda $. Suppose that $P(\lambda )$ is a solution of a given
equation of physics $F(\lambda ,P)=0$ that we are unable to solve exactly.
In some cases we can obtain an approximation to $P(\lambda )$ in the form of
a power series $P(\lambda )=P_{0}+P_{1}\lambda +\ldots $ by means of
perturbation theory. If this series is divergent or slowly convergent we may
try and improve the results by means of a resummation technique.

Variational perturbation theory consists of modifying the physical equation
in the form $F(\xi ,\alpha ,\lambda ,P)=0$, where $\alpha $ is a variational
parameter (or a set of them in a more general case) and $\xi $ is a dummy
perturbation parameter so that $F(1,\alpha ,\lambda ,P)=F(\lambda ,P)$.

Then we apply perturbation theory in the usual way, calculate $N+1$
coefficients of the perturbation series
\begin{equation}
P(\xi ,\alpha ,\lambda )=\sum_{j=0}^{\infty }P_{j}(\alpha ,\lambda )\xi ^{j}
\end{equation}
and construct an approximation of order $N$ to the property
\begin{equation}
P^{[N]}(\alpha ,\lambda )=\sum_{j=0}^{N}P_{j}(\alpha ,\lambda ).
\label{eq:P[N]}
\end{equation}
If the partial sums $P^{[N]}(\alpha ,\lambda )$ converged toward the actual
property as $N\rightarrow \infty $, then $P^{[\infty ]}(\alpha ,\lambda
)=P(\lambda )$ would be independent of $\alpha $. However, for finite $N$
the partial sums do depend on the variational parameter $\alpha $. It is
therefore reasonable to assume that the optimum value of this parameter
should be given by the principle of minimal sensitivity (PMS) \cite{S81}:
\begin{equation}
\left. \frac{\partial }{\partial \alpha }P^{[N]}(\alpha ,\lambda )\right|
_{\alpha =\alpha _{N}(\lambda )}=0.  \label{eq:PMS_P}
\end{equation}
In many cases $P^{[N]}(\alpha _{N}(\lambda ),\lambda )$ converges towards $%
P(\lambda )$ as $N\rightarrow \infty $, and, besides, $P^{[N]}(\alpha
_{N}(\lambda ),\lambda )$ behaves like $P(\lambda )$ with respect to $%
\lambda $ even at relatively small perturbation orders.

\bigskip

P.A. acknowledges support of Conacyt grant no. C01-40633/A-1.

\end{document}